\documentclass[%
 reprint,
 amsmath,amssymb,
 aps,
]{revtex4-2}

\usepackage{graphicx}
\usepackage{dcolumn}
\usepackage{bm}
\usepackage{siunitx}
\usepackage{subcaption}

\begin{document}

\preprint{APS/123-QED}

\title{Anomaly Detection at the European XFEL using a Parity Space based Method}

\author{Annika Eichler, Julien Branlard, Jan H. K. Timm}
\affiliation{%
Deutsches Elektronen-Synchrotron DESY, Germany
}%





\begin{abstract}
A novel approach to detect anomalies in superconducting radio-frequency cavities is presented, based  on  the  parity  space  method with the goal to detect quenches and distinguish them from other anomalies. The model-based parity space method relies on analytical redundancy and generates a residual signal computed from measurable RF waveforms. The residual is a sensitive indicator of deviation from the model and provides different signatures for different types of anomalies. This new method not only helps with detecting faults, but also provides a catalogue of unique signatures, based on the  detected  fault. The method was experimentally verified at the European X-ray Free Electron Laser (EuXFEL). Various types of anomalies  incorrectly detected  as  quenches  by  the  current  quench detection  system are analysed using this new approach.
\end{abstract}

\maketitle


\section{Introduction}
Large scale superconducting particle accelerators such as the European X-ray Free Electron Laser (EuXFEL) comprise several hundreds radio frequency (SRF) cavities. A user facility of this size expects high RF availability in order to accelerate beam and provide users with reliable and predictable photon light. Automation algorithms~\cite{JB_SRF2021} are necessary to monitor the RF and recover as fast as possible stations where a trip took place. One typical issue causing RF down time is a cavity quench, where an area of the cavity walls becomes normal conducting (thermal breakdown), which translates into a drop of the cavity quality factor by several order of magnitudes leading to a collapse of its accelerating field~\cite{padamsee1998rf}. The likelihood of a quench taking place increases when cavities are operated at high gradient, close to their quench limits (20 - 30 MV/m for EuXFEL cavities). For pulsed accelerators, the usual approach to quench detection consists of measuring the loaded quality factor $Q_L$ during the decay phase of the RF pulse~\cite{Branlard:ICALEPCS13-THPPC072}. While this approach has proven to be robust detecting actual quenches, it also provides false positive cases, when anomalies in machine conditions have an impact on the $Q_L$ computation. In its current implementation, the quench detection system cannot discriminate a real from a "fake" quench and provides by default the same reaction: switching off the RF. However, more information than what is currently used by the quench detection server is available in the control system, and can be used to help analyse each RF pulse, with the intent to discriminate real quenches from other anomalies. The current quench detection server uses the cavity probe amplitude information to compute $Q_L$. The probe phase, the RF forward and reflected signals are also available and can be used to provide a more accurate description of the type of fault taking place. The electrical and mechanical behaviour of SRF cavities is well understood and can be modelled~\cite{Schilcher1998}. Thus, model-based fault detection methods can be used to detect faults as it has been demonstrated in \cite{nawaz2018anomaly, Nawaz2021, Bellandi2021}. These model-based approaches provide a promising insight on anomaly detection and categorization. As an alternative, a fully data-based approach is also possible as presented in \cite{Tennant2020}, where machine learning is applied for the identification and classification of faults in SRF cavities for the Continuous Electron Beam Accelerator Facility (CEBAF) at Jefferson Laboratory. The basis for this approach was a data set of several thousand events, manually labeled. 

In addition to cavity failures, fault and anomaly detection has gained significant interest for accelerator operation in general due to the great potential to increase availability. Mainly data-driven methods have been used, exploiting tools from machine learning. Application examples range from magnet fault detection at CERN~\cite{Fol2019} and at the Advanced Photon Source at Argonne National Laboratory \cite{Edelen2021} over the detection of faulty beam position monitors at the Large Hadron Collider \cite{Wielgosz2017a} to fault detection at the digital electronics level \cite{Grunhagen2021,Martino2021}.

In this paper, a novel approach to detect cavity anomalies is presented based on the parity space method \cite{nawaz2018anomaly}. This method relies on the well-known cavity model, since no extensive labeled data-set is available at DESY for a purely data-driven approach. The fault detection parity method makes use of analytical redundancy to generate a residual signal from measurable RF waveforms. This residual can be further analysed with statistical tools to evaluate significant deviations from the model, i.e., a fault.  Resulting measures provide different signatures depending on the type of detected anomaly. This method can detect faults but also provides a catalogue of signatures based on the detected fault, thus allowing for a more robust cavity quench detection, by minimizing the number of false positives. The method was experimentally verified at EuXFEL and applied to various types of anomalies incorrectly detected as quenches by the current quench detection server. The next Section \ref{sec:cavities} provides general information about SRF cavities and their mathematical model. Different types of faults observed, based on four years of operation at EuXFEL are summarized in Section \ref{sec:cavityfaults}. The following Section \ref{sec:parity} provides an insight on the parity space method for fault detection and evaluation exploiting the cavity model.  Experimental results are given in Section \ref{sec:results}.

\section{SRF Cavities}\label{sec:cavities}
\begin{figure}
      \includegraphics[scale=0.35]{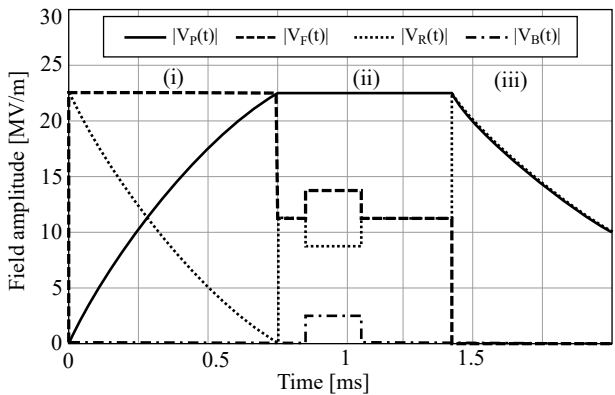}
      \caption{Amplitude of the fields for the probe $|V_P(t)|$, forward $|V_F(t)|$ and reflected $|V_R(t)|$ waveforms, and the field imposed by the beam  $|V_B(t)|$ for the three regions of an RF pulse: (i) filling, (ii) flattop and (iii) decay.}
      \label{fig:signals}
\end{figure}

SRF cavities are electromagnetic resonators that can be modelled as second order systems in the I (in-phase) and Q (quadrature) domain as \cite{Schilcher1998}
\begin{equation}
%
\begin{split}
\begin{bmatrix}
	\dot{V}_{P,I}(t)\\
	\dot{V}_{P,Q}(t)\\
\end{bmatrix} =& \begin{bmatrix}
	-\omega_{1/2} & -\Delta\omega(t) \\
	\Delta\omega(t) & -\omega_{1/2} \\
\end{bmatrix}  \begin{bmatrix}
	V_{P,I}(t)\\
	V_{P,Q}(t)\\
\end{bmatrix}	
\\
	&\: + \: 2\omega_{1/2}\begin{bmatrix}
	V_{F,I}(t)\\
	V_{F,Q}(t)\\
\end{bmatrix}
-\: \omega_{1/2}\begin{bmatrix}
	V_{B,I}(t)\\
	V_{B,Q}(t)\\
\end{bmatrix},
\end{split}
\label{eq:cavity}
\end{equation}
where $V_F(t) = V_{F,I}(t)+jV_{F,Q}(t)\in\mathbb{C}$ is the forward field coupled into the cavity. Here, $V_{F,I}(t)$ and $V_{F,Q}(t)$ are the I and Q components, respectively. The state $V_P(t) = V_{P,I}(t)+jV_{P,Q}(t)\in\mathbb{C}$ is the probe signal, i.e. the field which builds up inside the cavity, and $V_B(t)\in\mathbb{C}$ is the field induced by the beam. Note that all signals are considered in base-band. This means that $V_F(t)$ represents the envelope of the electromagnetic forward wave which is driven by the main oscillator with a frequency of $f=1.3$\,GHz. Currently, the European XFEL is operated in pulsed mode with an RF pulse repetition rate of \SI{10}{Hz}. Each pulse lasts for approximately \SI{1.8}{ms} and can be divided into filling, flattop and decay. During filling, the electromagnetic forward wave is coupled into the cavity so that the probe electromagnetic standing wave increases up to the desired field gradient. It is held constant via a feedback controller during the flattop, to accelerate the arriving electron beam by the desired energy level. In the decay, the forward electromagnetic wave coupled into the cavity is switched off and the probe wave decays. $V_P(t) = V_F(t) + V_R(t) + V_B(t)$ for all $t$. Figure~\ref{fig:signals} shows the amplitude of the cavity signals in the different operating regions. 

The parameters influencing the dynamic behavior of the SRF cavities are the half bandwidth $\omega_{1/2}$ and the detuning $\Delta \omega(t)$. The half bandwidth $\omega_{1/2}=\pi f/Q_L$ is defined by the loaded quality factor $Q_L$, which expresses the ratio of the energy stored inside the cavity to the dissipated energy. The external input power coupling to the cavity, defined by the penetration depth of the coupler antenna, is denoted as $Q_{ext}$. The unloaded, external and loaded quality factors are linked by the following relationship~\cite{padamsee1998rf}:
\begin{align}\label{eq:quality_factor}
    \frac{1}{Q_L} = \frac{1}{Q_0} + \frac{1}{Q_{ext}}\,.
\end{align}
Due to the superconductivity of the cavities at the European XFEL, the unloaded quality factor is very high ($\approx 10^{10}$). The external quality factor can be altered to tune the loaded quality factor. In nominal operation the European XFEL is operated at a quality factor around  $Q_L=4.6 \cdot 10^6$, resulting in a small half bandwidth ($\approx 140$~Hz). This can be easily estimated during the decay region of the RF pulse as one over the time constant of the exponential decay of the probe signal. 

Due to its small half bandwidth, the cavity is sensitive to the detuning $\Delta \omega(t)$, the second parameter that influences the cavity dynamics. The detuning is the difference between the driving frequency $f$ and the cavities resonance frequency $f_0(t)$. Ideally, this should be zero, but the resonance frequency is determined by the cavities shape, which changes over time as the cavity is exposed to forces, e.g., mechanical forces and Lorentz forces. In principle the detuning can be modelled as a second order mechanical resonator. However, due to the short duty-cycle an approximation as sum of first order systems is given in \cite{Schilcher1998} as

\begin{align}
	\Delta\dot{\omega}_n(t) =& -\frac{1}{\tau_n}\Delta\omega_n(t) + K_n(V_{P,I}^2(t) + V_{P,Q}^2(t))\, , \notag\\ & n =1,...,N \notag, \\
	\Delta\omega(t) =& \sum_{n=1}^{N}\Delta\omega_n(t)\,.
	\label{eq:detuning}
\end{align}
Here $N$ defines the number of relevant mechanical modes (index $n$), $\tau_n$ the time constant of the response of the respective modes and $K_n$ is the Lorentz force constant.

\section{Anomalies that triggered the current quench detection}\label{sec:cavityfaults}
The quench detection and reaction system deployed at EuXFEL~\cite{Branlard:ICALEPCS13-THPPC072} triggers if the $Q_L$ computed for a single pulse drops below the running average by more than a user-defined threshold (typically 10\% of the nominal $Q_L$ value). In this contribution, four particular conditions or anomalies yielding false positives on the quench detection system are investigated: a controlled cavity detuning, a controlled change of cavity bandwidth, an electron burst linked to a spontaneous field emission, and finally a digital glitch corrupting the data transfer at the firmware or software level. These four cases triggered the quench detection system, although no real quench occurred. 

\subsection{Controlled change of bandwidth}
Coupler heating (due to operation) can change the cavity external quality factor $Q_{ext}$ by up to 25\%~\cite{Cichalewski-TNS2019} for TESLA cavities. When increasing or decreasing the XFEL linac energy to accommodate different user run requirements, the change in cavity forward power is accompanied by a net change in $Q_{ext}$, due to heating effects. At EuXFEL, the input power couplers are motorized, allowing for remote cavity bandwidth adjustments. As a routine maintenance, the $Q_{ext}$ are monitored and readjusted. While the thermal effects are very slow (several hours), a motorized coupling readjustment can produce a rapid change in $Q_{ext}$. This change is detected as a change in $Q_L$ as it can be seen in Eq.~(\ref{eq:quality_factor}) by the quench detection system. In some conditions, if the change is fast enough, the quench detection will falsely interpret this change as a quench. Several countermeasures are possible: one can mask the quench detection system when adjusting $Q_{ext}$, or adjust the learning rate to low-pass filter these rapid controlled $Q_L$ variations. Finally, one could slow down the motorized coupler drive or increase the complexity of the quench detection system by having it check for motor movements. Having an independent measurement that can provide additional information to help in the algorithm decision process is the approach chosen in this contribution.

\subsection{Controlled detuning}
As reported in~\cite{Branlard:ICALEPCS13-THPPC072}, changing the cavity tuning has an influence on the measured cavity coupling. A \SI{1}{kHz} change in cavity resonance can infer a deviation of up to 15\% in cavity bandwidth at EuXFEL. This coupling comes from the measurement error due to the limited isolation between the forward and reflected ports of the waveguide bi-directional couplers. When detuning the cavity, the increased reflected traveling wave couples out to the incident signal, changing its amplitude and phase. While the cavity external quality factor, $Q_{ext}$, remains constant, the measured loaded quality factor $Q_L$ is affected. In some cases, a fast detuning (requested by operation for example) can fool the quench detection system and result in a station trip. Some safety mechanisms are built in to avoid such false positives: for example ignoring detected quenches if the cavity gradient is below a user-defined minimal threshold (\SI{2.5}{MV/m} typically). Heavily detuned cavities are hence ruled out, but this exception handling does not cover all cases. Mechanical coupling between the deformation of the cavity due to the tuner motor and the penetration of the coupler's antenna can also be observed. At EuXFEL however, the impact of this mechanical coupling is of second order compared to the electrical cross talk described above.


\subsection{Field emitter and other electronic processes}

Spontaneous field emission and multipacting~\cite{FARNSWORTH1934411}~\cite{Knobloch:SRF97-SRF97D23} can be observed sporadically on some cavities. During the cryomodule test phase~\cite{Branlard:SRF13-THP087}, prior to XFEL tunnel installation, cavities prone to field emission were identified and flagged. In some cases, their operating gradient was limited to keep the emitted radiation below the acceptance threshold~\cite{PhysRevAccelBeams.20.042004}. However, new field emitters can appear during operations, and can be triggered through secondary emission due to the onset of another field emitter. Such events produce erratic beam loading, often observed as flickering and noisy cavity RF traces taking place at random times during the RF pulse. In some extreme cases, the produced dark current is large enough to discharge the cavity field within tens of nanoseconds. This effect also referred to as plasma discharge can affect neighbouring cavities, that will see this beam loading, albeit in a reduced magnitude. If this disturbance takes place during the decay phase of the RF pulse, it will influence the computation of the $Q_L$ value, falsely triggering the quench detection system. These spontaneous electronic processes are typically accompanied by radiation bursts, observed by gamma and neutron detectors in the tunnel. In most cases, field emitters are conditioned away after a few pulses, but not always.  A simple conditioning attempt is not always successful or could result in generating new field emitters via secondary emission. Techniques such as plasma processing have been developed to cure field emission~\cite{Knobloch:1998zs} but are currently not implemented at DESY. 

\subsection{Digital glitches}
Several failures in the digital domain have been observed so far, resulting in a corruption of the data transferred between electronic boards. Due to the fact that all electronics responsible for cavity field control are located inside the tunnel, they are subject to radiation showers that can produce single event upsets, flipping a bit in the digital data stream. There are countermeasures in place to track and fix such events (check-sum or cyclic redundancy checks) but multiple bit flips taking place simultaneously cannot be fixed with this approach. Another failure mode has been observed when the CPUs in the tunnel become temporarily overloaded, delaying data memory accesses to the time when the FPGA is writing to memory. These read/write collisions  result in data loss, where not all data points are recorded. These two failure cases can result in discontinuities in the cavity waveforms, corrupting the $Q_L$ computation. Such events typically trigger the quench detection system, although no real quench occurred.

\section{Fault Detection}\label{sec:parity}
In the following section we present a method for fault detection, the parity space method. A model-based approach is chosen because it does not require large data sets for training, and since a good model description of the cavities is available, the method is robust against changes in operating point. With the parity space method a residual is generated, as described in Section \ref{subsec:resgen}, that can be further evaluated by statistical tests, see Section \ref{subsec:ResEval}. 

\subsection{Parity Space Method}
The parity method is a method for fault detection that is based on analytical redundancy. Analytical redundancy relations are derived from an analytical model, as the cavity model in \eqref{eq:cavity} and \eqref{eq:detuning}, and only involve measured variables \cite{kinnaert2003fault}. These have to hold in absence of a fault. Thus, the analytical model has to represent the nominal behavior of the system. The deviation from the analytical relation is called residual, i.e., if the residual is zero, the system behaviour is as described by the model, thus behaving nominally, otherwise it is behaving faultily.

Consider the  nonlinear system 
\begin{align}
    \dot{x}(t) & = f(x(t),u(t),d(t),\nu(t)) \label{eq:state},\\
    y(t) & = h(x(t),u(t),d(t),\nu(t))\label{eq:output}.
\end{align}
Here \eqref{eq:state} is called the state equation and $x(t)\in\mathbb{R}^n$ is the state vector at time $t$, $u(t)\in\mathbb{R}^m$ the control input, $d(t)\in\mathbb{R}^{m_d}$ is the disturbance vector, representing sensor noise, unmodelled dynamics, etc. The fault signals are described by $\nu(t)\in\mathbb{R}^{m_\nu}$. Equation \eqref{eq:output} is called the output equation and $y(t)\in\mathbb{R}^n$ is the output vector. For fault detection a residual $r(t)$ has to be described, depending on known signals only, that is zero in the absence of faults and different from zero otherwise \cite{bokor2009fault}, i.e., 
\begin{align}
    \left\{\begin{array}{lll}\nu(t)=0 &\Rightarrow & r(t)=0\\
    \nu(t)\neq 0 &\Rightarrow & r(t)\neq 0\\
    \end{array}\right.\quad \forall u(t),\,y(t),\,d(t).
    \label{eq:residual_cond}
\end{align}
It is clear that in real system the residual will never be exactly zero due to the disturbances $d(t)$, which can not be measured. Therefore, instead of \eqref{eq:residual_cond}, the residuals need to be robust against disturbances $d(t)$ but at the same time sensitive to faults $\nu(t)$. We will tackle this in the following with a stochastic interpretation of the residuals and introduce with the generalized likelihood a measure when the residual significantly deviates from zero, thus the behaviour is not consistent with the nominal behaviour, i.e., there is a fault. This will be presented in detail in Section \ref{subsec:ResEval}.

\subsection{Residual Generation}\label{subsec:resgen}
It is assumed in the following that the output $y(t)$, the input $u(t)$ as well as multiple derivations of these signals are known, but not the states $x(t)$ (and neither $d(t)$ nor $\nu(t)$).
To calculate the residuals, analytical redundant expressions of the model description are exploited. These can be generated by deriving the single equations of \eqref{eq:state} and \eqref{eq:output} multiple times and eliminating the unknown states to obtain residuals that are only dependent on the known signals 
\begin{align*}
    r_i = P_i(u,y,u^{(1)},u^{(2)},\ldots,u^{(\alpha_{i})},u^{(\alpha_{i})})\,,\quad i = 1,2,\ldots\,,
\end{align*}
where $i$ denotes the $i$-th residual. With $u^{(k)}$ the $k$-th derivative of $u(t)$ is defined. Depending on the residual $i$, the highest degree $\alpha_i$ of deviation might be different. While the elimination of the unknown states is obvious for linear systems \cite{kinnaert2003fault}, it can be very involved for nonlinear ones \cite{bokor2009fault}.

Given the model of SRF cavities  \eqref{eq:cavity} and \eqref{eq:detuning}, three (redundant) relations for the non-measurable detuning $\Delta \omega(t)$ can be derived as 
\begin{align*}
    &\Delta \omega_{I}(t) = \frac{-\dot{V}_{P,I}(t) \!+\! \omega_{1/2}\big(\!-\!{V}_{P,I}(t)\! +\! 2{V}_{F,I}(t)\! -\!V_{B,I}(t)\big)}{V_{P,Q}(t)}\,,\\
    &\Delta \omega_{II}(t) = \frac{\dot{V}_{P,Q}(t) \!+\! \omega_{1/2}\big({V}_{P,Q}(t) \!-\! 2{V}_{F,Q}(t) \!+\!V_{B,Q}(t)\big)}{V_{P,I}(t)}\,,\\
    &\Delta \dot{\omega}_{III}(t) = \sum_{n=1}^N \Delta \dot{\omega}_n(t)\,.
\end{align*}
While $\Delta \omega_{I}(t)$ and $\Delta \omega_{II}(t)$ are already functions of known signals only, $\Delta \omega_{III}(t)$ is a function of the non-measurable states $\Delta {\omega}_n(t)$ for $n=1,\ldots,N$. As the relationship is linear, standard procedures as in given in \cite{kinnaert2003fault} exist to eliminate the non-measurable states.
Then, three different residuals can be derived, i.e., $\Delta \omega_{I}(t)-\Delta \dot{\omega}_{II}(t)$, $\Delta \dot{\omega}_{I}(t)-\Delta \omega_{III}(t)$ and $\Delta \dot{\omega}_{II}(t)-\Delta \dot{\omega}_{III}(t)$. As the first has been shown in \cite{nawaz2018anomaly} to be most informative, we will concentrate on 
\begin{align} \label{eq:residual_cont}
    r(t) = & \; \Delta \omega_{I}(t)-\Delta \omega_{II}(t)\\
    = & \;\frac{-\dot{V}_{P,I}(t) + \omega_{1/2}\big(-{V}_{P,I}(t) + 2{V}_{F,I}(t) -V_{B,I}(t)\big)}{V_{P,Q}(t)} \notag\\ 
    & -  \frac{\dot{V}_{P,Q}(t) + \omega_{1/2}\big({V}_{P,Q}(t) - 2{V}_{F,Q}(t) +V_{B,Q}(t)\big)}{V_{P,I}(t)}\,. \notag
\end{align}
Further reasons for this choice are that (i) for this residual no additional derivation of the data are required, which amplifies noise, and (ii) $\Delta \omega_{I}(t)$ and $\Delta \omega_{II}(t)$ only depend on $\omega_{1/2}$, which is either known or can be easily determined, as described in Section \ref{sec:cavities}, while  $\Delta \omega_{III}(t)$ depends on $\tau_n$ and $K_n$ for $n=1,\ldots,N$, for which an additional system identification step is necessary. 

The continuous physical cavity system, $y(t)$ with $t\in\mathbb{R}$, is sampled in discrete time $y(t_0+kT)$ for $k\in\mathbb{Z}$, where $T$ is the sampling period. This will be abbreviated as $y(k)$ in the following. In order to account for this, the residual \eqref{eq:residual_cont} needs to be discretized. For the sampling rates in question (\SI{1}{MHz} or \SI{9}{MHz} depending on which layer the algorithm is implemented), Euler forward discretization can be used with $\dot{y}(t)\approx\frac{y(k+1)-y(k)}{T}$. Replacing $\dot{V}_{P,I}(t)$ and $\dot{V}_{P,Q}(t)$ accordingly in \eqref{eq:residual_cont} leads to 
\begin{align} \label{eq:residual_dis}
    r(k) = &\;\frac{1}{ V_{P,Q}(k)T}\bigg(-{V}_{P,I}(k+1) + {V}_{P,I}(k)  \notag\\ & \quad\quad+  \omega_{1/2}T\big(-{V}_{P,I}(k) + 2{V}_{F,I}(k) -V_{B,I}(k)\big)\bigg) \notag\\ 
    & -  \frac{1}{V_{P,I}(k)T} \bigg({V}_{P,Q}(k+1)-{V}_{P,Q}(k)  \notag\\ & \quad\quad + \omega_{1/2}T\big({V}_{P,Q}(k) - 2{V}_{F,Q}(k) +V_{B,Q}(k)\big)\bigg)\,.
\end{align}
To improve the numerical properties of this residual, for implementation \eqref{eq:residual_dis}  is multiplied with $V_{P,I}(k)V_{P,Q}(k)T$.

\subsection{Residual Evaluation}\label{subsec:ResEval}
There are several approaches to evaluate residuals; in this work we apply statistical tests. Here, likelihood ratios are indicators of the goodness of fit of a null hypothesis $H_0$ versus an alternative one $H_1$ by the ratio of their likelihoods, assessing if the observed residuals supports or significantly disagrees with the null hypothesis.

Assuming that the observed residuals follows a statistical model with parameter $\mu$, with $\mu=\mu_0$ being the null hypothesis and $\mu=\mu_1$ being the alternative one, the log-likelihood ratio given the observed residuals $r(k),\ldots,r(k+K-1)$ is defined as \cite{Ding2008}
\begin{align}
    \label{eq:likelihood}
    \lambda_{\text{LR}}(k) = \sum_{i=k-K+1}^{k}\ln \frac{p(\mu_1|r(i))}{p(\mu_0|r(i))}\,.
\end{align}
Here $p(\mu_1|r(k))$ is the probability of the alternative hypothesis $H_1$ given the observed residual $r(k)$; the complimentary probability $p(\mu_0|r(k))$ is defined accordingly. For the application at hand, we assume in case of a fault or anomaly, i.e., in case of the alternative hypothesis, a jump in the mean value of a Gaussian distribution as
\begin{align}
    \label{eq:gauss}
    {r}(k)  = \mu + \mathcal{N}(0,\,\Sigma), \text{ with } \theta = \!\begin{cases}
    \mu_0 = 0,&\!\!\! H_0(\text{no change})\,,\\
    \mu_1 \neq 0,& \!\!\!H_1(\text{change})\,.
   \end{cases}
\end{align}
With this in nominal operation, we expect that the residuals follow a zero-mean Gaussian distribution with variance $\Sigma$, while in case of a fault or anomaly in the system, a jump in the mean value appears. So that the data is still Gaussian distributed with the same variance but the mean value is different. While the variance $\Sigma$ can be calculated from the given nominal data, the mean $\mu_1$ is unknown. For estimating $\mu_1$,  the maximum log-likelihood ratio, i.e., generalized likelihood ratio (GLR), is considered, derived from  \eqref{eq:likelihood} by replacing  $\mu_1$ by its maximum likelihood estimate, given as \cite{Ding2008}
\begin{align}
    \label{eq:max_likelihood}
    \lambda_{\text{GLR}}(k) & = \max_{\mu_1} \lambda_{\text{LR}}(k)\\& = \frac{K}{2} \left(\frac{1}{K}\!\sum_{i = k-K+1}^{k}\!\!\!\!\! r(i)^\top\right) \Sigma^{-1} \,\left( \frac{1}{K}\!\sum_{i = k-K+1}^{k}\!\!\!\!\! r(i)\right)\,,\notag
\end{align}
Under the assumption \eqref{eq:gauss}, $2\lambda_{\text{GLR}}$ follows a $\chi^2(n)$ distribution. Here, $n$ is the dimension of the residuals, which in our case is one, since only the one-dimensional residual \eqref{eq:residual_cont} is considered. With this an error threshold $\bar{\lambda}_{\text{GLR}}$ can be chosen from the $\chi^2(1)$ distribution, so that $\lambda_{\text{GLR}}(k)>\bar{\lambda}_{\text{GLR}}$ is considered to be erroneous (or anomalous), according to a chosen acceptable probability of false positives alarms given as $P(Q>2 \bar{\lambda}_{\text{GLR}})$ with $Q\sim \chi^2(1)$.

In order to fulfill the assumption \eqref{eq:gauss} for the nominal case, the residuals coming from the parity space \eqref{eq:residual_cont} need to be zero mean. Thus, we  correct them for the mean value as
\begin{align}\label{eq:mean_correction}
    r'(k) = r(k)-\bar{r}(k)\quad \text{with}\quad \bar{r}(k)=\frac{1}{P}\sum_{p=1}^P r_p(k)\,,
\end{align}
where $\bar{r}(k)$ is the mean value of sample $k$ over $P$ pulses. It is clear that this would not be necessary if the model equations \eqref{eq:cavity} and \eqref{eq:detuning} would perfectly fit and \eqref{eq:gauss} would only contain the measurement noise. But as discussed above, this is never the case in reality. Thus repetitive disturbances or model mismatches can be corrected using Eq.~(\ref{eq:mean_correction}). This approach also allows taking the beam contribution into account. As obvious in model \eqref{eq:cavity} the beam is an input here. Although it can be measured, the beam information was not available in the given data sets. Thus, the beam loading is considered as a repetitive disturbance, whose effect is cancelled by the mean value correction in \eqref{eq:mean_correction}.

\section{Results}\label{sec:results}
In the following section, the parity space method for fault detection is applied to real operational data, and the different cases described in Section \ref{sec:cavityfaults}, that falsely trigger the quench detection system, are presented and analysed.
\subsection{Data}
The data that will be analysed in this section is provided by a snapshot recorder, triggered by any interlock occurring in the RF system. For each snapshot, the cavity signals of 250 pulses, i.e., 25 seconds, are saved, with 200 pulses before the event and 50 afterwards. Unfortunately, the beam signal is not saved, but a good workaround to deal with the missing signal is presented in Section~\ref{sec:implementation}. Each pulse in the data set is labeled with a unique pulse identifier, the ID, and the data also includes a time vector. The data is saved as hdf5 files. One data sets contains the forward, reflected and probe signals in amplitude and phase, which can be easily transferred to the I and Q domain. Time domain signals are \SI{1.82}{ms} long, sampled at \SI{1}{MHz} yielding 1820 samples per waveform.

\subsection{Implementation}\label{sec:implementation}
The fault detection infrastructure is implemented in C++. The analysis is automated with an easy-to-use console command. Although only results of a-posteriori analysis are presented here, the algorithm is optimized for real-time and can cope with \SI{10}{Hz} RF pulse repetition rate for pulse-to-pulse detection. This run-time operation capability was demonstrated for a small amount of cavities. See \cite{Timm2021} for further implementation details.

\subsection{Statistics}
From 07/08/2020 till 11/18/2020, 34 snapshots were saved, triggered by  the quench detection system. 
After a postmortem analysis of these trips, 18 of them have been identified as real quench. This yields a false positive rate of 47\%. Thus, almost half the events identified as quenches are no quenches during this period of operation. Within this period of 15 weeks, the European XFEL was operated at high gradient for five weeks (i.e. 16.5~GeV beam energy) and ten weeks at low gradient (14~GeV beam energy). Two third of the real quenches occurred within the weeks of high gradient. 

\begin{figure}
\begin{subfigure}[t]{0.45\linewidth}
\includegraphics{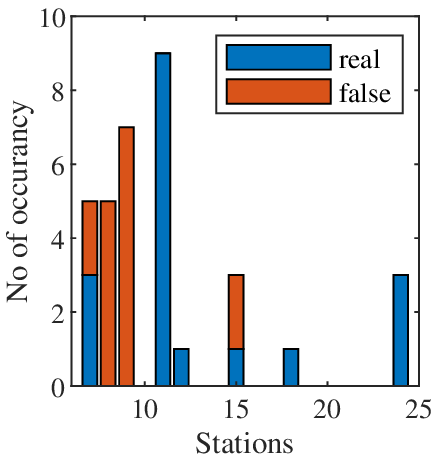}
\subcaption{Occurrences of real quenches and false positives per station}
\label{fig:station_dist}
\end{subfigure}%
\hfill%
\begin{subfigure}[t]{0.45\linewidth}
\includegraphics{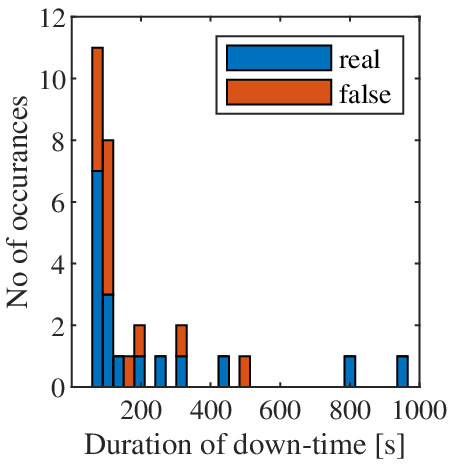}
\subcaption{Contribution of real quenches and false positives on down-time}
\label{fig:duration_dist}
\end{subfigure}
\caption{Statistics of archived events triggered by the quench detection and reaction system}
\end{figure}

Figure~\ref{fig:station_dist} shows the distribution of the events, i.e., real quenches and false positive, over the different stations (denoted as A2 to A25). Station A11 is the highest performing station and was operating very close to its quench gradient during the high-gradient study period. This explains the higher quench rate. Otherwise, the quenches are relatively spread over the stations. The duration of down-time caused by the respective event is shown in Fig.~\ref{fig:duration_dist}. Here one real quench with a down-time of more than an hour is missing. Due to this outlier, the mean down-time for the real quenches is almost seven minutes, while it is two minutes for the false positives. The median however is very similar in both cases around \SI{100}{s}.

\begin{figure}
\begin{subfigure}[t]{1\linewidth}
\includegraphics{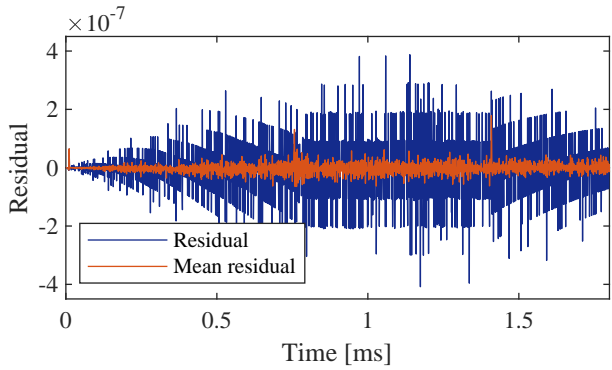}
\subcaption{Without beam}
\end{subfigure}%
\\
\begin{subfigure}[t]{1\linewidth}
\includegraphics{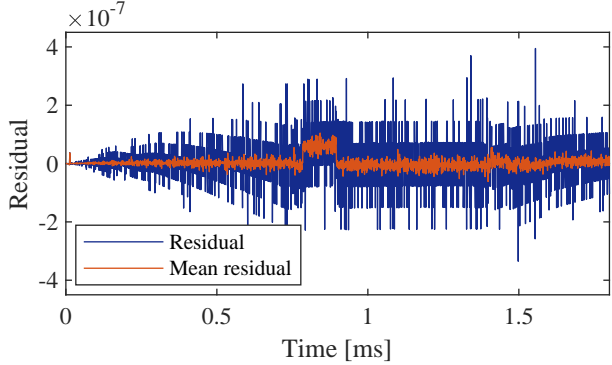}
\subcaption{With beam}
\end{subfigure}
\caption{Nominal residuals without and with beam. In blue the residuals of one pulse are shown, in red the residuals are averaged over 50 pulses. }
\label{fig:residuals}
\end{figure}

\subsection{Residual Generation and Evaluation}
\begin{figure}
\begin{subfigure}[t]{0.45\linewidth}
\includegraphics{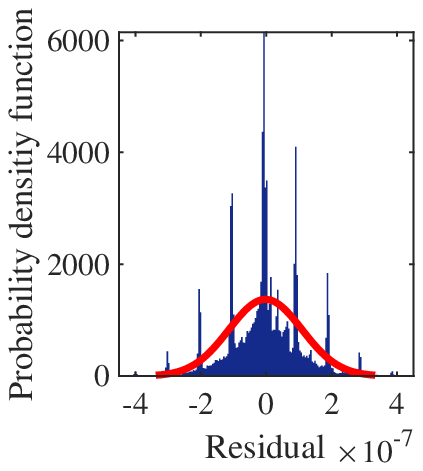}
\subcaption{Histogram if residuals}
\label{fig:histogram}
\end{subfigure}%
\hfill%
\begin{subfigure}[t]{0.45\linewidth}
\includegraphics{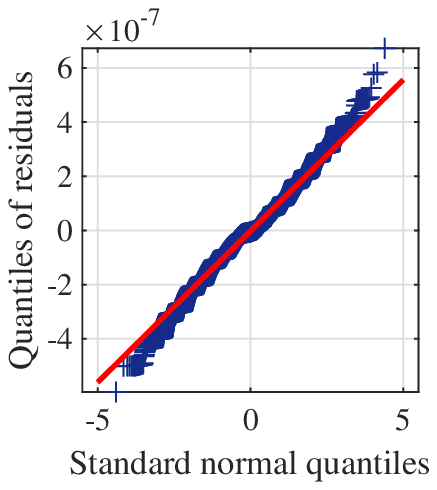}
\subcaption{Q-Q plot of residuals}
\label{fig:qqplot}
\end{subfigure}
\caption{Probability plots of residuals without beam. In red a fitted normal distribution is shown.}
\end{figure}

The residuals are calculated as given in \eqref{eq:residual_dis}, scaled with $V_{P,I}(k)V_{P,Q}(k)T$ for numerical issues as discussed in Section~\ref{subsec:resgen}. Figure~\ref{fig:residuals} shows the residuals of nominal pulses. The averaged residuals without beam show that the model fits very well and the residuals are mean value free. Since the presence of beam is not yet implemented as an input to the model and not available as measurements in the data, the presence of accelerated beam is clearly visible as a model mismatch as shown in Fig.~\ref{fig:residuals}. As the beam signal is not changing during the short snapshot data sets that we consider, we can correct for this using~\eqref{eq:mean_correction}. Furthermore, it is obvious that with the residual calculation we are in the resolution range of the considered signals as one clearly sees quantization effects. This is confirmed in the histogram in Fig.~\ref{fig:histogram}. As the Q-Q plot in Fig.~\ref{fig:qqplot} appears to be relatively linear, there might be an underlying Gaussian distribution, however this is clearly  distorted by quantization. With this, the assumption in  \eqref{eq:gauss} does not hold. Nevertheless, the GLR can be used as metric, as it can be interpreted as a weighted
error norm and the experimental distribution of nominal data can be considered to choose for a reasonable threshold. The upper bound  $\bar{\lambda}_{\text{GLR}}=10.8$, is chosen empirically, such that the given nominal data would have not led to a false positive alarm. This bound would have led to a false-positive rate of 0.0003\% if the ${\lambda}_{\text{GLR}}$ followed the chi2 distribution.

\subsection{Analysis of the Generalized Likelihood Ratio}

\subsubsection{Quenches}
\begin{figure}
\includegraphics{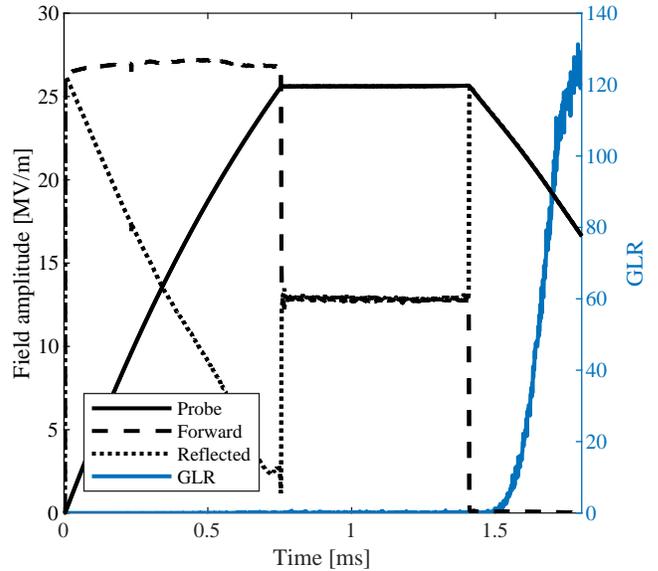}
\caption{\label{fig:quench1} Quench signals for ID 794495576.}
\end{figure}

\begin{figure}
\includegraphics{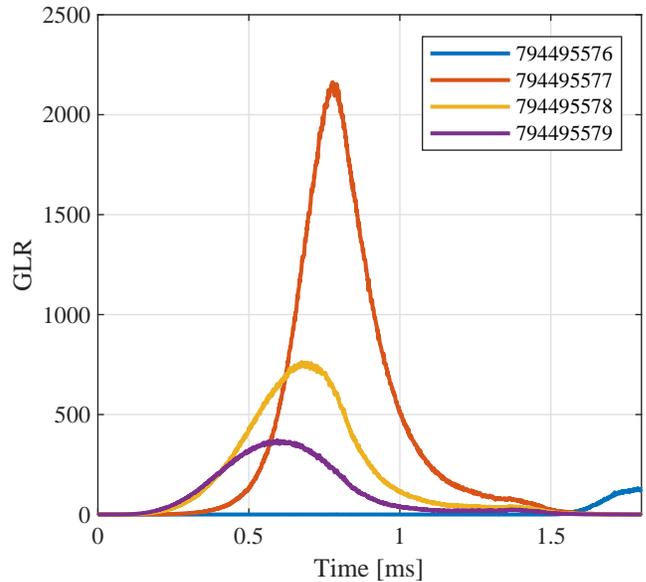}
\caption{\label{fig:quench2} GLR values of quenches for IDs 794495576-794495579.}
\end{figure}

Figure~\ref{fig:quench1} shows the cavity signals of a quench event together with the GLR for the first pulse, where the threshold of $\bar{\lambda}_{\text{GLR}}=10.8$ is exceeded. The GLR for this pulse and the three consecutive ones are shown in Fig.~\ref{fig:quench2}. The RF was switched off on the following pulse by the quench detection system. As shown here, there can be a delay of multiple pulses between quench detection and reaction, due to the current software implementation of the quench detection. Thus, it is unclear at what pulse the quench detection system has actually been triggered. All signals of the GLR follow a characteristic bell curve, initiating during the decay of the first pulse and occurring at earlier times for each successive pulse.
One quench event is selected here for demonstration but all of them show very similar GLR traces.

\subsubsection{Field emitter and other electronic processes}
\begin{figure}
\includegraphics{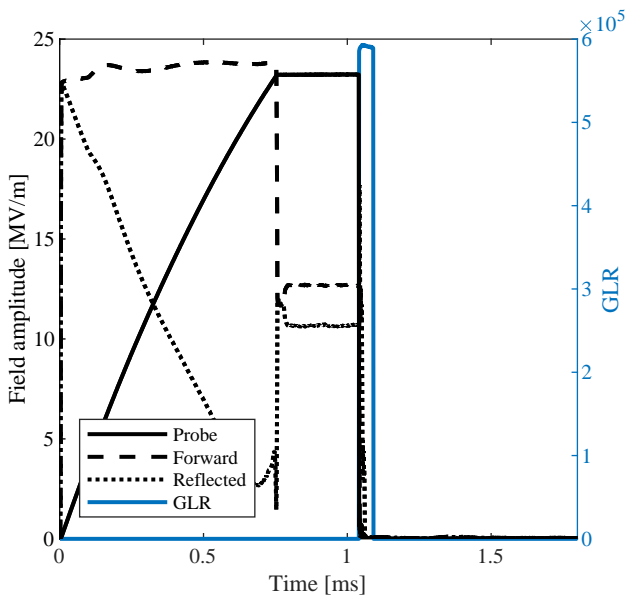}
\caption{\label{fig:multi1} Field emitter signals for cavity C4.M3., ID 1209385508.}
\end{figure}

\begin{figure}
\includegraphics{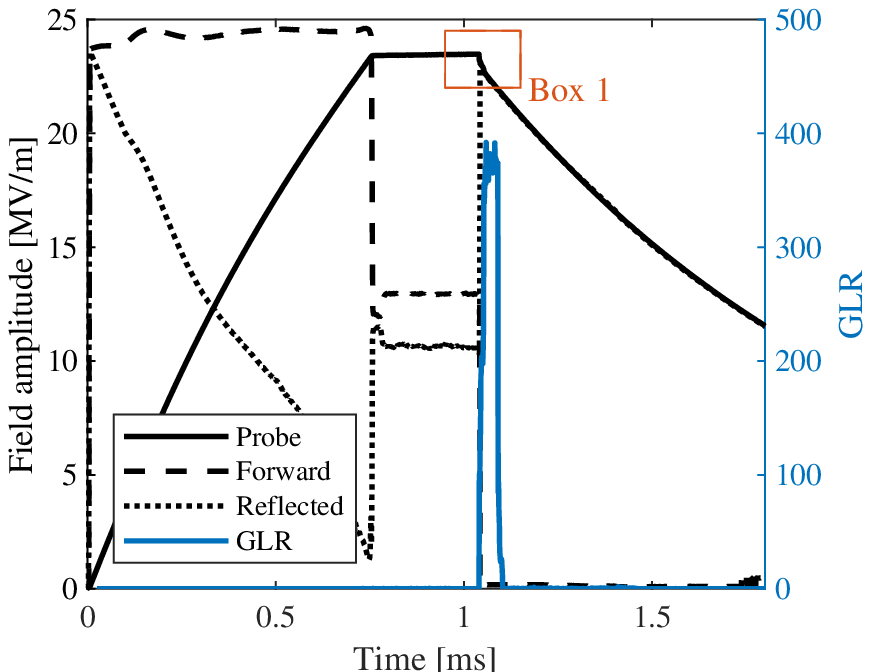}
\caption{\label{fig:multi2} Field emitter signals for cavity C1.M3, ID 1209385508.}
\end{figure}

\begin{figure}
\includegraphics{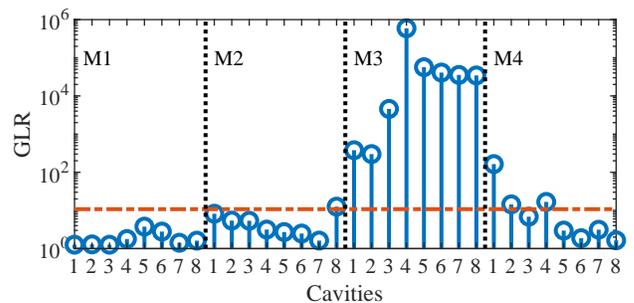}
\caption{\label{fig:multi3} Maximum GLR signal over the pulse  ID 1209385508 of all 32 cavities for the plasma discharge event observed on C4.M3. The threshold of $\bar{\lambda}_{\text{GLR}}=10.8$ is marked in red, dashed-dotted.}
\end{figure}

The RF signals observed during a plasma discharge event and the corresponding GLR for cavity 4 of cryomodule 3 (C4.M3) are shown in Fig.~\ref{fig:multi1}, where the accelerating field is discharged within tens of nanoseconds. This strong anomaly is confirmed by an extremely high GLR value. For this event, neighbouring cavities have also seen this beam loading effect, albeit in a reduced magnitude. Figure~\ref{fig:multi2} shows the discharge of a fraction of the probe field (see Box 1) for a neighboring cavity within the same cryomodule (C1.M3). The propagating effect downstream, but also upstream is further illustrated in Fig.~\ref{fig:multi3} where the maximal GLR value over the whole pulse is plotted for all cavities within the station. The cavity where the event took place (C4.M3) is clearly identified, its GLR being one order of magnitude higher than all others. One can further see the slow decay of the effect on downstream cavities (C5 to C8 in M3 and C1 to C4 in M4) and upstream (C3 to C1 in M3 and C8 in M2), where the threshold of $\bar{\lambda}_{\text{GLR}}=10.8$, marked in red is exceeded. Most of the energy is scattered at the quadruple located at the end of each cryomodule, due to the mismatch between the quadrupole setting and the energy of the accelerated dark current.


The GLR signature is clearly distinct from that observed in case of quenches. It can be further noticed for example in Fig.~\ref{fig:multi2} that very small effects in the cavity signals get strongly emphasized by the GLR.


\subsubsection{Glitch}
\begin{figure}
\includegraphics{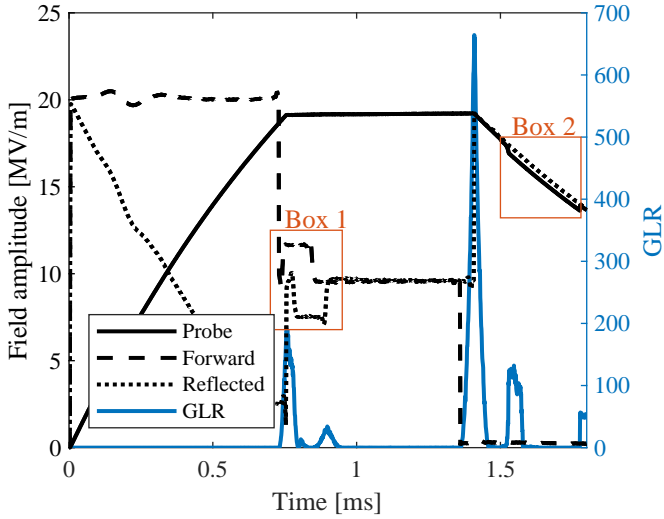}
\caption{\label{fig:glitch} Glitch signals.}
\end{figure}

The cavity signals for a digital glitch together with the GLR are shown in Fig.~\ref{fig:glitch}. Here the signals of one cavity are shown as an example, but all cavities within the module show the same distortions. It is obvious that this cannot be a physical fault, since forward and reflected signals are shifted in time during the flattop region, one compared to the other (see Box 1), while the probe and reflected waveforms differ during the decay (see Box 2). This is not physically possible, as the probe equals the sum of the forward and reflected signal, and the forward signal is zero during the decay.

\subsubsection{$Q_L$ Change}
\begin{figure}
\includegraphics{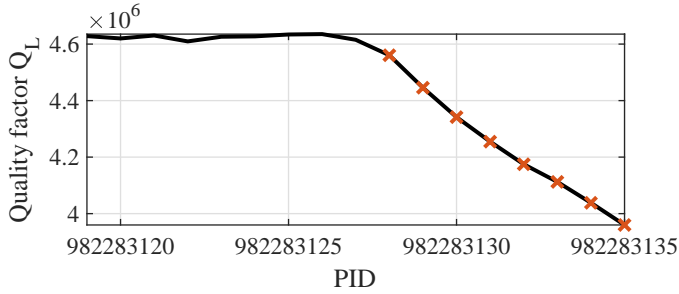}
\caption{\label{fig:Quality} Loaded quality factor $Q_L$ for IDs 982283119-982283135.}
\end{figure}

\begin{figure}
\includegraphics{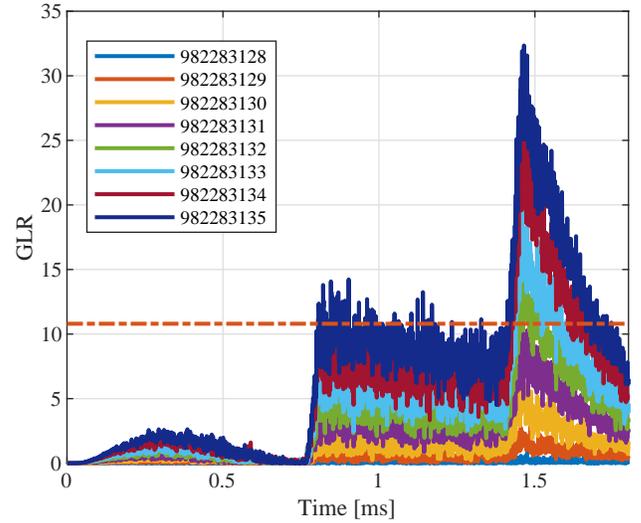}
\caption{\label{fig:QLall} GLR values for changed $Q_L$ for IDs 982283128-982283135.  The threshold of $\bar{\lambda}_{\text{GLR}}=10.8$ is marked in red, dashed-dotted.}
\end{figure}

\begin{figure}
\includegraphics{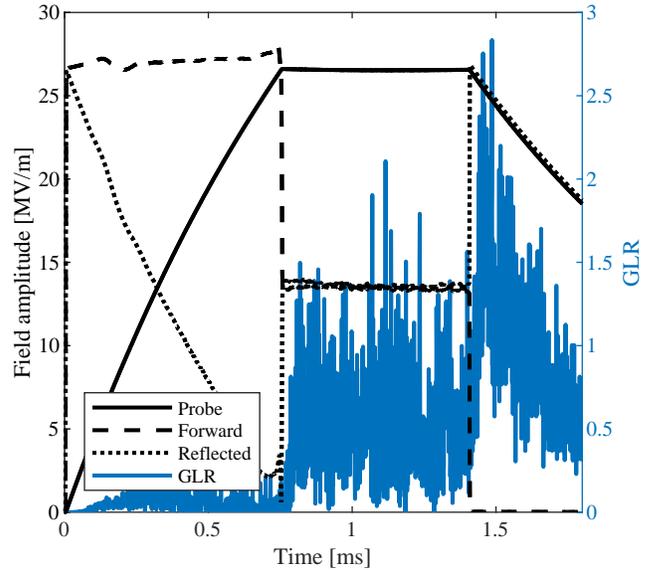}
\caption{\label{fig:QL1} All signals for changed $Q_L$ for ID 982283129. The threshold of $\bar{\lambda}_{\text{GLR}}=10.8$ is not hit.}
\end{figure}

\begin{figure}
\includegraphics{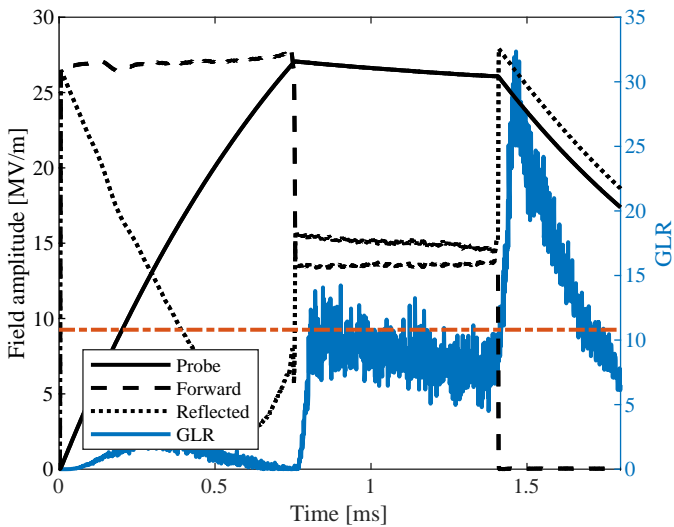}
\caption{\label{fig:QL2} All signals for changed $Q_L$ for ID 982283135. The threshold of $\bar{\lambda}_{\text{GLR}}=10.8$ is marked in red, dashed-dotted.}
\end{figure}

To study the effect of a controlled change in $Q_L$, the motorized antenna of an input power coupler was deliberately moved over several pulses. This change in $Q_{ext}$ directly impacts the measured $Q_L$, as shown in Fig.~\ref{fig:Quality}. Figure~\ref{fig:QLall} shows the GLRs corresponding to the eight successive pulses marked by red crosses in Fig.~\ref{fig:Quality}. While the first one looks normal (numerical noise), a clear signature is visible starting from the second pulse, remaining relatively flat in the filling, followed by a first step up when the flattop begins and a second one at the beginning of the decay followed by a linear decay. The amplitude of this specific signature increases as $Q_{ext}$ is further reduced. Two specific pulses (the second and the last one of those depicted in Fig.~\ref{fig:QLall}) are shown in detail in Fig.~\ref{fig:QL1} and \ref{fig:QL2}. While the threshold is not hit for the second pulse in Fig.~\ref{fig:QL1}, one clearly sees the distinct signature. The threshold is hit two pulses later. After the eighth pulse, shown in detail in Fig.~\ref{fig:QL2}, the quench detection system has switched off the cavity.

\subsubsection{Detuning}
\begin{figure}
\includegraphics{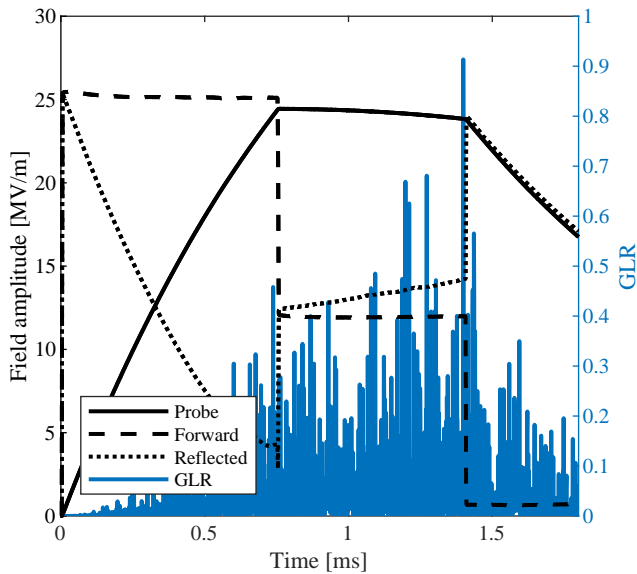}
\caption{\label{fig:detuning1}All signals for changed for changed detuning for ID 764126695.}
\end{figure}

\begin{figure}
\includegraphics{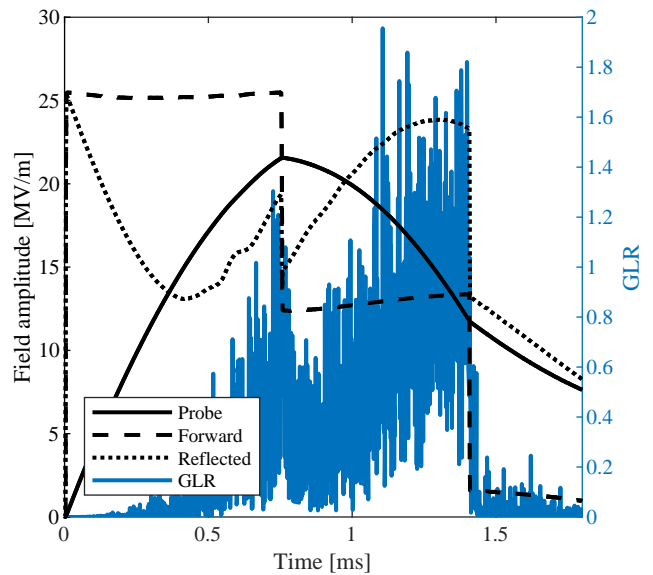}
\caption{\label{fig:detuning2} All signals for changed for changed detuning for ID 764126702.}
\end{figure}

Figures~\ref{fig:detuning1} and~\ref{fig:detuning2} show an event, were the quench detection system is fooled by a fast detuning change. The first change in cavity detuning is observed in Fig.~\ref{fig:detuning1} (positive tilt of the reflected signal during the flattop region). Larger detuning is evidenced in Fig.~\ref{fig:detuning2}, characterized by the round drop of the cavity probe gradient and steeper increase of the reflected signal during the flattop. Also here the GLR shows a special signature, distinct from the ones observed before: a linear increase during the filling and the flattop while it is almost zero during the decay. It is interesting to note that the GLR stays below the threshold  despite the heavy detuning. This is to be expected, as the residual in \eqref{eq:residual_cont} is the difference between detuning values resulting from two differential equations of the same electromagnetic model \eqref{eq:cavity}. If we assume that the model holds, a common mode detuning change will affect both equations identically, hence yielding a zero residual. The reason that we see distinct traces can be explained by model imperfections, but they do not significantly affect the fault detection, as it is desired here not to fool the detection.


\section{Conclusion}
This work presents the parity space method for fault detection in SRF cavities followed by statistical evaluation using the generalized likelihood ratio. It validates the good performance of the method in detecting faults and also shows that different fault types result in clearly distinct signatures of the generalized likelihood ratio. The distinction of different fault types could also be achieved by a thorough analysis of all six cavity signals (I and Q values of probe, forward and beam signals). However, with the generalized likelihood ratio, all information is embedded into one single signal. This is of practical use for online operation, as the operator only needs to observe one signal but also facilitates the a-posteriori classification of faults. Furthermore, this opens the door towards automatic classification of faults using classification tools from machine learning, which is subject to future work. Such an online automatic classification would be of great support for linac operation. 

It could be demonstrated that, due to the choice of residuals, the method presented here is robust against changes in cavity detuning. Thus, in contrast to the current quench detection system, the proposed approach would not result in false positives of this kind. Detecting such abrupt detuning changes could be realized using additional residuals. One possibility would be to solve the electromagnetic model for the half bandwidth. It is expected that this should be independent of a change of $Q_L$, but detect a change in the detuning.

In this work, the cavity faults have been analysed post mortem. However, the code for fault detection and evaluation is able to cope with real-time requirements and can be switched to support offline and online analysis. The online analysis has already been demonstrated for a small amount of cavities. It will be subject to future work to bring this to normal operation for all cavities. 

\section{Acknowledgments}
The authors wish to thank Sven Pfeiffer and Denis Kostin for fruitful discussions about the data analysis during the writing of this paper.

\bibliography{ParitySpaceResidual}

\end{document}